\pdfoutput=1
\documentclass[twocolumn,secnumarabic,amssymb, nobibnotes, aps, pra]{svjour3}

\newcommand{\macro}[1]{\texttt{\textbackslash#1}}

\usepackage{amstext}
\usepackage{multirow}
\usepackage[logonly]{trace}
\usepackage[final]{graphicx}
\usepackage{booktabs}
\usepackage{rotating}
\usepackage{amsmath}
\usepackage{mathptmx,amssymb}
\usepackage{latexsym}
\usepackage{graphicx}

\RequirePackage{fix-cm}

\begin{document}
\smartqed  
\newcommand{\SJour}{\textsc{SVJour}}
\title{Reducing the effect of thermal noise in optical cavities}
\author{ Sana Amairi        \and
Thomas Legero \and
Thomas Kessler \and
Uwe Sterr \and
        Jannes B. W\"ubbena \and
        Olaf Mandel   \and
        Piet O. Schmidt}
\institute{Sana Amairi, Thomas Kessler, Jannes B. W\"ubbena, Olaf Mandel, Piet O. Schmidt  \at
              QUEST Institute, Physikalisch-Technische Bundesanstalt, 38116 Braunschweig, Germany \\
              Tel.: +49(0)531 592 4704\\
              \email{sana.amairi@quantummetrology.de}           
           \and
           Thomas Legero, Uwe Sterr \at
              Physikalisch-Technische Bundesanstalt, 38116 Braunschweig, Germany \\
           \and
           Piet O. Schmidt \at
            Institut f\"ur Quantenoptik, Leibniz Universit\"at Hannover, 30167 Hannover, Germany\\
                 \email{Piet.Schmidt@quantummetrology.de} }
\maketitle
\begin{abstract}
Thermal noise in optical cavities imposes a severe limitation in the stability of the most advanced frequency standards at a level of a few $10^{-16}\sqrt{\mathrm{s}/\tau}$ for long averaging times $\tau$. In this paper we describe two schemes for reducing the effect of thermal noise in a reference cavity. In the first approach, we investigate the potential and limitations of operating the cavity close to instability, where the beam diameter on the mirrors becomes large. Our analysis shows that even a 10~cm short cavity can achieve a thermal noise limited fractional frequency instability in the low $10^{-16}$ regime. In the second approach, we increase the length of the optical cavity. We show that a 39.5~cm long cavity has the potential for a fractional frequency instability even below $10^{-16}$, while it seems feasible to achieve a reduced sensitivity of $<10^{-10}$/g for vibration-induced fractional length changes in all three directions.
\end{abstract}

\section{Introduction}

Optical cavities with spectrally narrow resonances play an important role in metrological applications, such as optical frequency standards~\cite{Nicholson2012comparison,gill_optical_2005,chou2010frequency,rosenband_frequency_2008,ludlow_sr_2008}, gravitational wave detection~\cite{hough_search_2005} and tests of fundamental physics~\cite{braxmaier2001proposed,herrmann_rotating_2009}. In a typical setup for optical frequency standards, a clock laser is stabilized to one of the modes of a linear Fabry-P\'{e}rot (FP) cavity consisting of two mirrors separated by a spacer. Several noise sources, such as pressure and temperature fluctuations, but also vibrations, change the optical path length of the cavity. These disturbances can be reduced by placing the cavity in a vacuum chamber provided with the proper active and passive thermal control and careful isolation from mechanical and acoustic vibrations. During the past years, significant effort was put into designing reference cavity systems with inherent insensitivity to vibrations~\cite{nazarova_vibration-insensitive_2006,chen2006vibration,tao2007decreased,webster_vibration_2007,guyomarch_aspects_2009,millo_ultrastable_2009,dawkins_ultra-stable_2009,dube2009narrow,zhao_vibration-insensitive_2009,leibrandt_spherical_2011,webster_force-insensitive_2011}.
On a more fundamental level, thermal noise induces optical path length fluctuations \cite{gillespie_thermally_1995}. Brownian motion causes local random displacement in the cavity spacer, mirror substrates and mirror coatings, limiting the achievable length stability of optical reference cavities. A theoretical description based on the Fluctuation Dissipation Theorem (FDT) indicates that the thermal noise strongly depends on material parameters, such as the mechanical loss angle ($\phi$), but also on the size of the optical mode on the mirrors~\cite{levin1998internal}. One of the dominant sources for thermal noise arises from the mirror coatings due to their high mechanical loss~\cite{numata2004thermal,hong2012brownian}. Increasing the mode size of the beam on the mirror surface significantly reduces the corresponding thermal noise contribution~\cite{numata2002direct}. This can for example be achieved through the excitation of higher-order transverse modes \cite{mours_thermal_2006}.

Here, we analyze the feasibility of two alternative approaches to increase the mode size on the mirrors. The first approach is based on  cavities operated close to instability, i.e. with a near-planar or near-concentric mirror configuration~\cite{siegman_lasers_1986}, whereas the second relies on long cavities. In the latter approach, relative frequency fluctuations are further suppressed since they scale with the inverse of the cavity length.
In this paper, we show through simulations that a 10~cm long cavity, when operated near instability, can achieve a thermal-noise-limited instability of $1.5\times 10^{-16}$ in 1~s, whereas a 39.5~cm long cavity has the potential to achieve an instability below $10^{-16}$ in 1~s. A major challenge for long cavities is the required insensitivity to accelerations. We present a 39.5~cm long cavity design with a vibration sensitivity of the fractional length change of $<10^{-10}$/g in all three directions, assuming realistic machining tolerances.
We start in Section~\ref{sec:thermalnoise} with a short overview of the analytical model of the thermal noise and we give the results of finite element simulations of the frequency fluctuations in short and long cavities, with stable and near-unstable configurations.
Possible technical limitations of near-concentric and near-planar cavities arising from the dense mode structure close to instability, and an estimation of the effects on the laser frequency stabilization are investigated in Section~\ref{sec:near stability cavities}. Section~\ref{sec:mounting} focuses on the design of a vibration insensitive 39.5~cm long cavity.


\section{Thermal noise of optical reference cavities}
\label{sec:thermalnoise}
The fluctuation dissipation theorem (FDT) relates the thermal noise to unavoidable mechanical losses in the system~\cite{callen1951irreversibility,callen1952theorem}. The resulting relative displacement noise amplitude $\sqrt{S_L}/L$ is converted into fractional frequency noise by the relation:
\begin{eqnarray}
\label{eq:frequency_fluctuation}
     \sqrt{S_L}/L = \sqrt {S_\nu}/\nu.
\end{eqnarray}
Numata \textit{et al.} derived analytical expressions for the thermal noise contributions of different parts of the cavity system~\cite{numata2004thermal}, that have been further improved by Kessler \textit{et al.}~\cite{kessler_thermal_2012}.
The power spectral density of the displacement noise in a spacer of length $L$ and mechanical loss $\phi_{\mathrm{spacer}}$ can be written as:
\begin{eqnarray}
\label{eq:spacer}
     S_\mathrm{spacer}(\omega)=\frac{4k_BT}{\omega}\frac{L}{A~E}\phi_\mathrm{spacer},
\end{eqnarray}
where $k_B$ is the Boltzmann constant, $\omega$ is the angular noise frequency, $T$ is the temperature, $A$ is the end face area of the spacer excluding the center bore, and $E$ is Young's modulus. The frequency noise from thermal noise of the spacer scales as $1/\sqrt{L}$ as shown in Eq.~\ref{eq:frequency_fluctuation} and~\ref{eq:spacer}.
The thermal noise from the mirror substrate can be derived by modelling it as an infinite half space, resulting in the following expression:
\begin{eqnarray}
\label{eq:substrate}
  S_\mathrm{sub}(\omega)=\frac{4k_BT}{\omega}\frac{1-\sigma^{2}}{\sqrt{\pi}E}\frac{\phi_\mathrm{sub}}{w_0},
\end{eqnarray}
where $\sigma$ is Poisson's ratio, $w_0$ is the radius of the laser mode on the mirror, and $\phi_\mathrm{sub}$ is the mechanical loss of the substrate.
For a coating with thickness $d$, the simplified expression of its thermal fluctuation spectrum is:
\begin{eqnarray}
\label{eq:coating}
  S_\mathrm{coat}(\omega)=\frac{4k_BT}{\omega}\frac{2(1+\sigma)(1-2\sigma)}{\pi E}\frac{d}{w_{0}^2} \phi_\mathrm{coat},
\end{eqnarray}
where $\phi_\mathrm{coat}$ denotes the mechanical losses of a homogenous coating layer.

The dominant source of thermal noise for typical reference cavities arises from the mirror substrates and their coating ~\cite{numata2004thermal}. Choosing  fused silica for the mirror material, with its one order of magnitude higher mechanical quality factor $Q=1/\phi$ compared to ultra-low expansion glass, significantly reduces the thermal noise. Cavities made entirely of high mechanical $Q$ materials, such as single-crystal silicon \cite{kessler2012sub}, have been investigated.
This leaves the coatings on the cavity mirrors as the major contribution to the thermal noise owing to their more than two orders of magnitude larger mechanical loss compared to fused silica.
The analytical expression in Eq.~\ref{eq:coating} suggests several options to reduce the coating thermal noise that have been experimentally investigated.
Decreasing the mechanical loss $\phi_\mathrm{coat}$ by replacing the commonly employed amorphous SiO$_2$/Ta$_2$O$_5$ Bragg-reflector stacks by single-crystal Al$_x$Ga$_{1-x}$As coatings is a promising approach~\cite{cole_monocrystalline_2008}. Other proposals include thermal noise compensated multi-layer coating designs \cite{kimble_optical_2008,gorodetsky_thermal_2008}. Khalili suggested to reduce the thickness $d$ of the coating and compensate the loss in reflectivity by adding another mirror behind the first \cite{khalili_reducing_2005,somiya_reduction_2011}. The two separated coating stacks form an etalon, which results in high reflectivity of the overall two-mirror system when tuned into anti-resonance, whereas the thermal noise is dominated by the first (thinner) coating, leading to a lower thermal noise compared to a single-mirror high-reflective coating. Coating-free mirrors based on micro-structured surfaces are yet another possibility \cite{bruckner2010realization}. However, the low experimentally achieved optical reflectivity and the difficulty of producing mirrors with a curvature, as required for stable optical cavities, currently limits the use for optical frequency metrology.

According to Eq.~\ref{eq:substrate}~and~\ref{eq:coating}, the thermal noise contribution of the substrate and the coating scale with $1/w_0$ and $1/w_0^2$, respectively, which suggests to enlarge the TEM$_{00}$ mode on the mirror surface. The radius of the optical mode on the mirror surfaces is a function of the distance between the mirrors and their radii of curvature. Two-mirror optical cavities with radii of curvature $R_1$ and $R_2$ are optically stable for $0\leq g_1 g_2\leq 1$, where $g_i=1-L/R_i$~\cite{siegman_lasers_1986}. Close to instability, the mode sizes on the mirrors diverge. The two symmetric configurations close to instability are the near-planar ($R\gg L$, $g_i\approx 1$) and the near-concentric ($R \approx L/2$, $g_i\approx -1$) configuration. Most experimentally realized stable cavity designs use a flat mirror and a mirror with 0.5 to 1~m radius of curvature. This configuration is well within the stability regime and the mode radius on the mirrors increases with cavity length. Moreover, according to Eq.~\ref{eq:frequency_fluctuation}, the fractional frequency noise decreases while increasing the length of the cavity as $1/\sqrt{L}$ for the spacer contribution (See Eq.~\ref{eq:spacer}) and $1/L$ for the substrate and the coating contributions (Eq.~\ref{eq:substrate}~and~\ref{eq:coating}). Consequently, the best thermal-noise-limited fractional instabilities of $2\times 10^{-16}$ and $1\times 10^{-16}$ were achieved for cavities 29~cm~\cite{jiang_making_2011} and 40~cm~\cite{swallows2012operating} long, respectively. Both approaches, cavities close to instability and long cavities, pose different technical difficulties that will be addressed in sections~\ref{sec:near stability cavities} and~\ref{sec:mounting}.
In the following, we present numerical estimates of the thermal noise for various cavity geometries, in particular short near-planar cavities and long optical cavities, and compare them to existing systems. The calculations are based on a direct application of the FDT using Levin's theorem~\cite{levin1998internal}, implemented by Finite Element Simulations (FEM) using the commercial program Comsol Multiphysics~\cite{comsol_multiphysics_2010}. The simulation is similar to the work presented in \cite{kessler_thermal_2012}, except for the coating contribution to the thermal noise that we calculated using the analytical expression of Eq.~\ref{eq:coating}.

Table~\ref{tb:thermal_noise} provides a summary of the estimated Allan deviation $\sigma_\nu$ of the frequency (flicker) noise for 3 different geometrical cavity designs and lists the individual contributions from the spacer, the substrate, and the coating to the frequency instability.
The relative frequency instability is calculated for the implemented or planned operating wavelength of the respective cavity. However, the selected wavelength plays only a minor role with longer wavelengths providing a slightly better performance due to a larger mode size on the mirrors. For cavities with coating-dominated thermal noise, doubling the wavelength also doubles the thickness of the coating \cite{lalezari_private_2010}, leaving the overall noise contribution constant.


\begin{sidewaystable*}
\center
\begin{tabular}{ccccccccc}
\hline
\multicolumn{7}{r}{~~~~~~~Frequency noise $(Hz/\sqrt{Hz})$}\\
\multicolumn{7}{r}{Relative contribution ($\%$)} \\
\cline{5-7}
Case&Substrate/Spacer&$R_1$/$R_2$& $w_0$ & Spacer &Substrate&Coating& $\sigma_y$&References to similar\\
~&~&~&($\mu$m)& ~&~&~&($10^{-16}$)&geometries \\
\hline \\
\textbf{\emph{24~cm long cavities}, $\lambda$=563~nm, d = 3~$\mu$m} & ~ & ~ & ~&~ & ~&~ &~ &~  \\
\hline \\

A&ULE~/~ULE& 0.5m/$\infty$  &293/212 & 0.021& 0.11 & 0.061 & 2.97 & ~\cite{young_visible_1999,numata2004thermal} \\
~&~& ~ & ~ & 2~\% & 77~\% & 21~\%& ~ &~ \\

B&FS~/~ULE &0.5m /$\infty$  &293/212 & 0.021 & 0.027 & 0.058 & 1.51 &~ \\
~& ~&~ &~ & 9 ~\%& 17 ~\% & 74 ~\%& ~ & ~  \\
\hline \\

\textbf{\emph{10~cm long cavities}, $\lambda$=1064~nm, d = 5.3~$\mu$m} & ~ & ~ & ~ &~ & ~&~ &~ &~  \\
\hline \\

C&FS~/~ULE&0.5m/$\infty$  &260/291& 0.023& 0.033& 0.087&  4.05  &~\cite{millo_ultrastable_2009,webster2008thermal}\\
~&~& ~ & ~ & 6~\%& 12~\% & 82~\%& ~ &~ \\

D&FS~/~ULE &30m/$\infty$ &767/766 & 0.023 & 0.020 & 0.031 & 1.85 &~\\
~& ~ &~&~& 28~\%& 21~\%& 51~\%& ~&~ \\
\hline \\

\textbf{\emph{39.5~cm long cavities}, $\lambda$=1069~nm, d = 6~$\mu$m} &~&~&  ~&~& ~&~ &~ &~ \\
\hline \\

E&FS~/~ULE&1m/$\infty$  &524/408& 0.007 & 0.0064& 0.014 & 0.72 &~ \\
~&~ & ~ &~ & 17 ~\%& 14 ~\% & 69 ~\%& ~ &~ \\

F&FS~/~ULE&0.2m/0.2m &778/778 & 0.007& 0.0049 & 0.0083 & 0.50 &~\\
~&~ & ~&~ & 35~\%& 17~\% & 48~\%& ~&~ \\
\bottomrule
\end{tabular}
\caption{Frequency noise calculation for different materials and cavities. The cavities have a cylindrical (A-D) or rectangular shape (E-F) operating with a laser at wavelength $\lambda$. The diameters of spacers of case (A-B), (C-D) and (E-F) are 150~mm, 100~mm and 74~mm respectively. The diameter of the spacer has a negligible influence on the total thermal noise. The optical mode radius $w_0$ at the position of the mirror is a function of L and the mirror radius of curvature $R_{1,2}$. The relative frequency noise contribution of each component of the cavity is presented in addition to the calculated value of the total frequency noise $\sqrt{S_\nu(1\mathrm{Hz})}$. $\sigma_\nu$ is the fractional frequency instability with $\sigma_y = \sqrt{S_\nu(1\mathrm{Hz})}\sqrt{2(\ln 2)}/\nu$. For all geometries, the central bore in the spacer is 1~cm diameter, the temperature is $T=293$~K, $\phi_\mathrm{ULE}=1.6\times10^{-5}$, $\phi_\mathrm{FS}=10^{-6}$. The coating on each mirror is of thickness d and $\phi_\mathrm{coat}=4\times 10^{-4}$. We assume for ULE (Corning) a Young's modulus of $67.6$~GPa and a Poisson ratio of $0.17$~\cite{gulati1997ule}, and for FS a Young's modulus of $73.1$~GPa and a Poisson ratio of $0.17$~\cite{FusedSilica}.}
\label{tb:thermal_noise}
\end{sidewaystable*}

According to the table, the major contribution to thermal noise for all cases stems from the coated mirrors, which is in good agreement with the findings of  reference~\cite{numata2004thermal} and the experimentally achieved thermal noise limits of $3\times 10^{-16}$~\cite{young_visible_1999} and $6.7 \times 10^{-16}$ ~\cite{millo_ultrastable_2009} for (A) and (C), respectively.
For the cavities with 24~cm length, replacing the ULE mirror substrates of case (A) by lower mechanical loss FS substrates would improve the frequency stability by a factor of 2 (case (B)).

The thermal-noise-limited performance of 10~cm long cavities is treated with plane/concave (C) and near-planar (D) mirror configurations. In case (D) the mode radius on the mirrors is increased to 766~$\mu$m, resulting in a reduction of the frequency instability by a factor 2 compared to (C). It is remarkable that using an optical configuration near instability for a cavity of only 10~cm length provides a frequency stability comparable to the performance of a much longer cavity such as the 24~cm long cavity of case (B). This design is particularly attractive for applications requiring por\-table optical cavities~\cite{argence2012prototype,leibrandt_field-test_2011,vogt_demonstration_2011,webster_force-insensitive_2011}. Possible technical issues in the realization of such a cavity are discussed in section~\ref{sec:near stability cavities}.

Cases (E) and (F) compare the improvement in frequency instability for a 39.5~cm long cavity when moving from a plane/concave (E) to a near-concentric (F) configuration. The long cavities result in improvements in the frequency stability by factors of 4 and 6 compared to case (A),
while the improvement between the design of the stable plano-concave cavity (E) and the near-concentric cavity (F) is about 30~\%. Recently, a cavity similar to case (E) has been implemented with a slightly different mirror configuration, demonstrating a thermal-noise-limited instability of $1 \times 10^{-16}$~\cite{swallows2012operating}. Both cavity geometries close to instability offer superior ther\-mal-noise-limited performance compared to more stable geometries.

\section{Cavities with large mode field diameters}
\label{sec:near stability cavities}
Operating optical cavities close to instability results in an increased alignment sensitivity and a small higher order mode spacing.
The frequency spacing of higher-order $(m,n)$ Hermite-Gaussian modes \-TEM$_{mnq}$ of a particular longitudinal mode $q$ is given by the resonance condition~\cite{anderson1984alignment}:
\begin{eqnarray}
\label{eq:resonances}
\frac{\omega_{mnq}}{\Delta\omega_\mathrm{FSR}}=q+\frac{1}{\pi}(m+n+1)\arccos\left(\sqrt{g_1 g_2}\right),
\end{eqnarray}
where $\Delta\omega_\mathrm{FSR} = 2\pi\times c/(2L)$ is the free spectral range of the cavity with $c$ denoting the speed of light. The close spacing of higher-order modes can disturb the stabilization scheme of the resonance frequency. Furthermore, the cavity becomes sensitive to misalignment of the mirrors that can significantly shift the position and angle of the optical mode axis inside the cavity. This leads to an enhanced sensitivity to seismic and acoustic vibrations (see Section \ref{sec:mounting}).

For the remainder of this section, we consider cavities with higher-order mode spacing $\Delta\omega_{hom}=|\omega_{00q}-\omega_{01q}|\approx 2\pi\times 27$~MHz, corresponding to the cases (D) and (F) of Table~\ref{tb:thermal_noise}.

\subsection{Influence of higher-order modes on the error signal}

\begin{figure}
\center
  \includegraphics[width=0.5\textwidth]{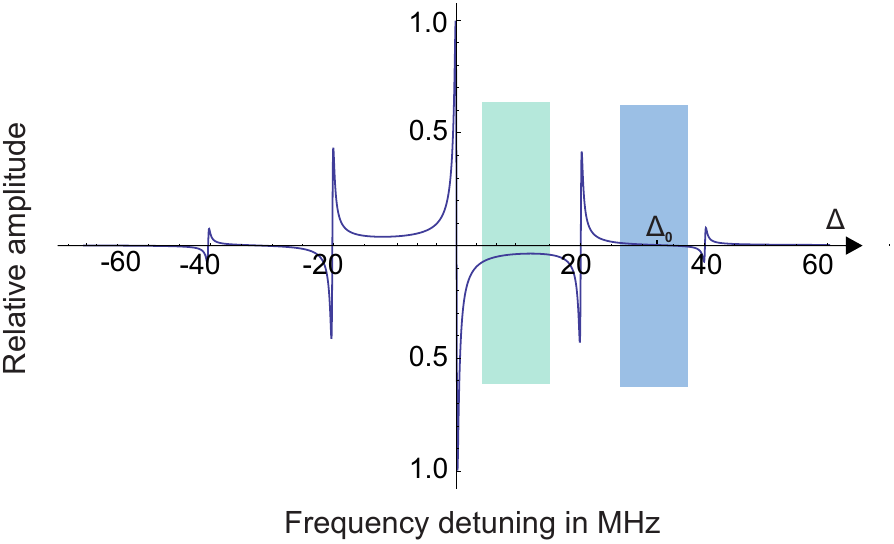}\\
  \caption{Error signal including first and second order modulation sidebands. The modulation frequency is $\Omega = 2\pi\times 20$~MHz, the modulation depth $\beta = 1$ and we assume for illustrative purposes a finesse of 5000. The green and blue regions are discussed in the text.}\label{Fig_errorsig_mod}
\end{figure}

The stability of passive optical cavities is typically transferred to a laser frequency using the Pound Drever Hall (PDH) technique~\cite{drever_laser_1983}. It is based on a measurement of the dispersive response of the cavity as a function of laser detuning from cavity resonance. In this scheme, sidebands are modulated onto the optical carrier at a frequency ranging between a few and a few hundred MHz. An example of the obtained error signal is shown in Fig.~\ref{Fig_errorsig_mod}. Typically, higher-order modes are neglected in the error signal analysis. However, imperfect mode matching into the cavity can lead to non-negligible coupling to these modes. Off-resonant coupling of the laser to these modes can result in a small offset in the error signal of the fundamental mode. Beam pointing fluctuations will then lead to a fluctuation in the offset of the error signal and therefore instabilities in the laser frequency.

\begin{figure}
\center
  \includegraphics[width=0.5\textwidth]{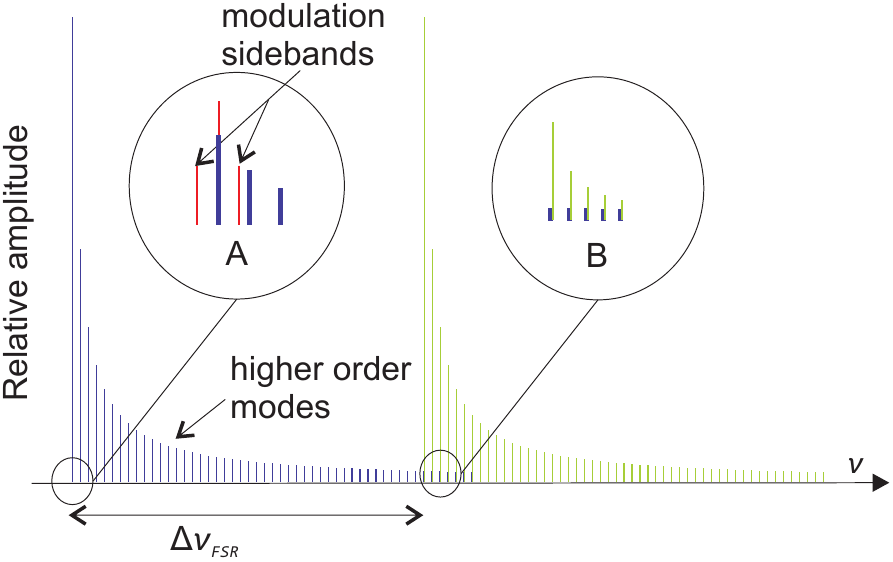}\\
  \caption{Example of two longitudinal modes (in green and blue) with higher-order modes distribution in a Fabry-Perot cavity of free spectral range $\Delta \omega_\mathrm{FSR} = 2\pi\times 1.5$~GHz. For illustrative purposes, the relative amplitude of the TEM$_{mnq}$ modes is scaled according to $1/(m+n+1)$. The circle (A) depicts the interaction of higher-order modes with a spacing $\Delta\omega_{hom}= 2\pi\times 27$~MHz with the frequency modulation sidebands at $\Omega =2 \pi \times 20$~MHz. The circle (B) illustrates the near-coincidence of the higher-order modes from a different longitudinal mode with the fundamental mode.}\label{Fig_coincidence}
\end{figure}

We investigated the two scenarios shown in Fig.~\ref{Fig_coincidence}: (A) Off-resonant coupling to the first higher-order mode near the fundamental resonance, and (B) near-coincidence of the fundamental resonance $\omega_{00q}$ with higher-order modes $\omega_{nmq \pm 1}$ of a different longitudinal mode of the cavity.
By carefully choosing the length of the cavity, the $\omega_{00q}$-resonance comes to lie close to the center between two higher-order modes. The required length accuracy of the cavity spacer of a cavity near instability $(g_i\approx \pm 1)$ can be estimated through
\begin{eqnarray}
\label{eq:coin}
 \delta L \ll \frac{{\Delta\omega_{hom}}^2}{\Delta\omega_\mathrm{FSR}\times \xi} = \Delta L_{hom},
\end{eqnarray}
where $\Delta L_{hom}$ is the required change in length to shift the resonance frequency $\omega_{00q}$ between two neighboring higher order modes and $\xi = \frac{\partial \Delta\omega_{hom}}{\partial L}$.
For $\delta L = \Delta L_{hom}/10$, we estimate that $\delta L$ for the 10~cm near-planar cavity [Table~\ref{tb:thermal_noise}(D)] is in the order of $370~\mu$m.
This result is compared to the case of the near-concentric 39.5~cm long cavity [Table~\ref{tb:thermal_noise}(F)] which has a similar higher order mode spacing and $\delta L$ is on the order of $70~\mu$m. However, for the same cavity finesse, the long cavity has a narrower linewidth compared to the short one, which relaxes the tolerances on $\delta L$. For the short cavity, coincidences occur at very large mode indices due to its large free spectral range, for which the coupling efficiency is expected to be small.

The effect of an off-resonant higher-order mode on the PDH error signal (case (A) of Fig.~\ref{Fig_coincidence}), is evaluated assuming a coupling efficiency $\eta_{hom}=0.1$ of the laser beam to this first higher-order mode of the cavity. In addition, we assume this coupling to fluctuate by $\delta \eta_{hom}/\eta_{hom}$ = 10\% to estimate the frequency fluctuations. This represents a worst case scenario. In the PDH setup, the error signal and thus the frequency shift are derived from a measurement of the power of the reflected phase-modulated beam.
For a lossless cavity, the reflection function $R(\Delta)$ of an incident beam with amplitude $\widetilde{E}(\Delta)$ is an Airy function that can be approximated by a Lorentzian:
\begin{eqnarray}
\label{eq:Rref}
  R(\Delta)=-\frac{\Delta (\Delta + \imath \Gamma/2)}{(\Gamma/2)^2+\Delta^2},
\end{eqnarray}
where $\Delta = \omega - \omega_0$ is the frequency shift of the laser of frequency $\omega$ from the nearest longitudinal mode $\omega_0$ and $\Gamma$ is the cavity linewidth.
The expression of the photodiode current signal is well known~\cite{riehle2004frequency}, and we expand it, taking into account the interaction between reflected fields up to the second order modulation sidebands.
The error signal is obtained through demodulation with a local oscillator field E$_0\sim\sin(\Omega t+\varphi )$ and low-pass filtering. For the usual situation of a fast modulation frequency ($ \Omega \gg \Gamma $) the error signal becomes:
\begin{equation}
\label{eq:errorsig}
\begin{split}
\varepsilon (\Delta) & = - 2 \emph{J}_0(\beta) \emph{J}_1(\beta)\Im [R(\Delta)R^*(\Delta+\Omega)\\
&-R^*(\Delta)R(\Delta-\Omega)]\\
&-2 \emph{J}_1(\beta) \emph{J}_2(\beta) \Im[R(\Delta+\Omega)R^*(\Delta+2\Omega)\\
&-R^*(\Delta-\Omega)R(\Delta-2\Omega)].
\end{split}
\end{equation}
Here, $\emph{J}_n(\beta)$ is the n$^{th}$ order Bessel function of the first kind as a function of the modulation index $\beta$, $\Im$ refers to the imaginary part, and $\Omega$ is the modulation frequency.

A plot of the normalized error signal is shown in Fig.~\ref{Fig_errorsig_mod}. Close to the cavity resonance ($\Delta \ll \Gamma$), the second order terms in $\Omega$ in Eq.~\ref{eq:errorsig} are negligible and $[R(\Delta)R^*(\Delta+\Omega)-R^*(\Delta)R(\Delta-\Omega)]\approx -\imath 2 \Im[R(\Delta)]$. Near resonance, we can approximate the error signal with a linear function of the frequency shift $\delta\omega$, where the slope is called the frequency discriminant $\textrm{D}$, and write the error signal near resonance $\varepsilon_{\mathrm{NR}}= \textrm{D} \times \delta\omega$ with:
\begin{equation}
\label{eq:slope_sig}
\begin{split}
\textrm{D} = - 8 \emph{J}_0(\beta) \emph{J}_1(\beta) \frac{1}{\Gamma}.
\end{split}
\end{equation}

The general expression of the frequency shift is then obtained by dividing the offset contributed by the off resonant higher order mode at detuning $\Delta$, $\delta\eta_{hom} \cdot \varepsilon_{hom} (\Delta)$ by the frequency discriminant $\textrm{D}$:
\begin{equation}
\label{eq:shift}
\begin{split}
\delta\omega(\Delta)= \delta\eta_{hom} \cdot \frac{\varepsilon_{hom}(\Delta)}{\textrm{D}}
\end{split}
\end{equation}

To understand the effect of the higher-order mode, e.g. TEM$_{010}$, on the PDH error signal, we consider 2 possible cases. The first case is for $\Gamma\ll\Delta\omega_{hom}< \Omega$ as shown in Fig.~\ref{Fig_errorsig_mod} by the green area around $\Delta \approx \Omega/2$.
As an example, we use Eq.~\ref{eq:shift} to calculate the laser frequency deviation $\delta\omega (\frac{\Omega}{2})$ due to a distortion of the error signal by a higher-order mode situated at $\Delta = \frac{\Omega}{2}$ for a 10~cm long cavity with a finesse $\mathcal{F}=10^5$, a modulation index of $\beta=1$, and a modulation frequency of $\Omega= 2 \pi \times 20$~MHz. In this case, the frequency fluctuation is estimated to be on the order of $\delta\omega (\frac{\Omega}{2}) = 2\pi \times 75 $~mHz. This value is too large for operating a laser with $<10^{-16}$ instability and therefore the first higher order mode resonance needs to be beyond the first modulation sideband.
The second case is for $\Omega < \Delta\omega_{hom}< 2~\Omega$ as shown in Fig.~\ref{Fig_errorsig_mod} by the blue colored area around $\Delta \approx 3\Omega/2$.
As an example, for a higher-order mode situated exactly at $3\Omega/2$ and for the parameters stated above, we get $\delta\omega(\frac{3\Omega}{2})=-2\pi \times 6.6 $~mHz, still representing a fairly large shift. However, the error signal has a zero crossing at $\Delta=\Delta_0 \approx 1.67~\Omega$ for our parameters.
If the first higher-order mode is situated at $\Delta=\Delta_0$ from the longitudinal mode resonance, this mode will have no influence on the error signal of the fundamental mode. This situation can always be achieved by adjusting the sideband modulation frequency. We evaluate the influence of the next higher order mode on the error signal for this case to be tolerable~$- 2 \pi \times 1.6$~mHz.

\subsection{Alignment tolerances}
\label{sec:Alignment tolerances}
\begin{figure}[h]
  \center
  \includegraphics[width=0.5\textwidth]{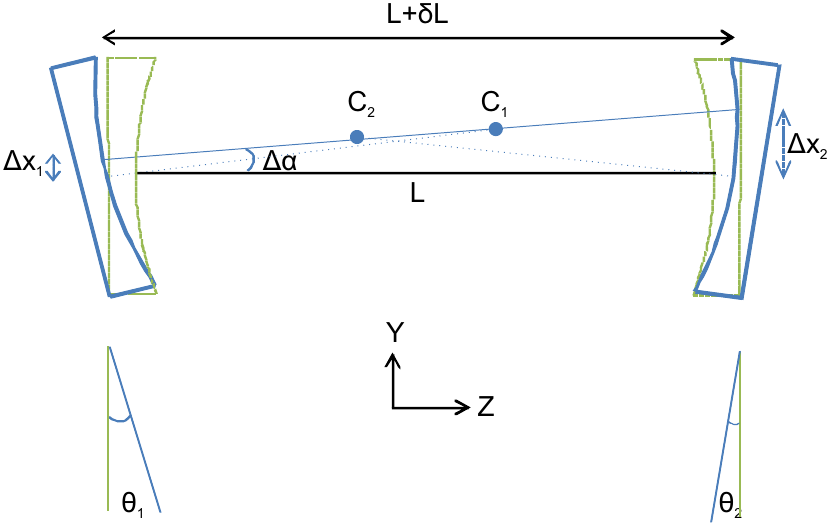}\\
  \caption{Illustration of the misalignment and the mode displacement in a stable FP resonator. The original positions are
drawn in green and the misaligned positions are drawn in blue.}\label{fig:mirror_tilt}
\end{figure}

Cavities close to instability are more sensitive to misalignment of the mirrors. This requires tight tolerances to spacer and mirror manufacturing, and to mirror alignment. The displacement of the optical mode from the center of the mirror affects the vibration sensitivity of the optical cavity as we will discuss in Section~\ref{sec:mounting}. In order to estimate the required tolerances, we consider a FP resonator made of two mirrors $S_{i}$ with the corresponding radii of curvature $R_i$ and centers $C_{i}$, separated by the optical path length of the cavity $L$ (see Fig.~\ref{fig:mirror_tilt}). If the mirror $S_i$ is tilted by an angle $ \theta_i$, the center of the mode intensity pattern on each mirror $S_i$ will shift due to a rotation of the optical axis by an angle $\Delta \alpha$ and a translation by a distance $\Delta x_i$ from the geometrical axis~\cite{siegman_lasers_1986} with:
\begin{eqnarray*}
\label{eq:alignement_equation}
    \Delta \alpha &=& \frac{(1-g_2)\theta_1-(1-g_1)\theta_2}{1-g_1g_2}\\
    \Delta x_1&=&\frac{g_2}{1-g_1g_2}\times L \theta_1+\frac{1}{1-g_1g_2}\times L \theta_2 \\
    \Delta x_2&=&\frac{1}{1-g_1g_2}\times L \theta_1+\frac{g_1}{1-g_1g_2}\times L \theta_2. \\
\end{eqnarray*}
For the case of a 10~cm long FP cavity with a near-planar mirror configuration [Table~\ref{tb:thermal_noise}(D)], we estimate a shift of $|\Delta x_1+\Delta x_2 |=100~\mu$m for a mirror tilt of $\pm 0.8~\mu$rad while with a near-concentric mirror configuration (R$_1$=R$_2$=0.51~cm), the same shift occurs for a mirror tilt of $\pm 0.9$~mrad.
We note that the sensitivity to mirror tilt scales with the length of the cavity for fixed g$_1$ and g$_2$.
In addition, for near-concentric cavities, the positioning of the mirrors on the end face of the spacer is critical, particulary for long near-concentric cavities. A shift $\Delta r$ of one mirror center from the symmetry axis of the cavity results in the rotation of the optical axis by an angle $\gamma \simeq \frac{\Delta r}{2R-L}$. This leads to a displacement $\Delta x = \gamma \times R$ of the optical mode on both mirrors. We estimate for a 39.5~cm long cavity [Table~\ref{tb:thermal_noise}(F)], that a shift of 10~$\mu$m of one mirror results in a mode displacement of 400~$\mu$m.
We conclude from these estimates that for a long cavity, working with a configuration close to instability requires challenging mechanical machining and alignment tolerances. However, for a short cavity the required tolerances are technically feasible.

\section{Vibration insensitive long cavity}
\label{sec:mounting}
As indicated in Table~\ref{tb:thermal_noise}, long cavities can reach a thermal noise limited instability below $10^{-16}$. However, low-frequency seismic and acoustic noise in the environment of the cavity cause fluctuations of the optical path length, particularly for long cavities with their intrinsically higher sensitivity to vibrations.
Careful positioning of the mounting points allows to support the cavity spacer with significantly reduced sensitivity to accelerations in all three directions.
Although length fluctuation and bending of the cavity can in principle be compensated, imperfections of the spacer manufacturing and moun\-ting process lead to a non-zero vibration sensitivity.

We begin by presenting the analysis of the sensitivity to vibrations and the results for an optimized mounting configuration, before comparing different diameter-length ratio configurations of a 39.5~cm long cavity. We end the section by estimating the effect of asymmetric spacer machining and unequal force distribution on the vibration insensitivity of the cavity.

\subsection{Optimized design of a vibration insensitive long optical cavity}\label{sec:ultrastable_cavity}

Small acceleration forces acting on an elastic body such as a cavity spacer, lead to a linear deformation of its shape through Hooke's law.
The first order relative change of the optical path length in a cavity subject to accelerations $a_j$ in the direction $j$ is given by:
\begin{eqnarray}
\label{eq:vibrationsens}
\frac{\Delta L}{L}=-\frac{\Delta f}{f}=\sum_{j=x,y,z} a_j k_j+\sum_{j=x,y,z} a_j \kappa_j \Delta r .
\end{eqnarray}
The first sum on the right-hand side of Eq.~\ref{eq:vibrationsens} describes a relative length change under $a_j$ with the sensitivity coefficient $k_j$. The second sum denotes a relative change in optical path length due to a tilt of the mirrors characterized by the coefficient $\kappa_j$ in case the optical mode is displaced from the mechanical symmetry axis of the cavity by an average distance $\Delta r$. In the ideal case, where $\Delta r=0$, the mirror tilt has a second order effect on the length change ($\delta L$ in Fig.~\ref{fig:mirror_tilt}), which can be neglected~\cite{nazarova_vibration-insensitive_2006}.

We have performed simulations using the Finite Element Method (FEM)~\cite{comsol_multiphysics_2010}, to determine the optimum position of the supporting points for a 39.5~cm long cavity with a square cross section equal to 74~mm side length. The choice of this square cross section will be justified after comparing the cavity performance with different side lengths at the end of this section. For the simulation we assume that the ULE spacer material is perfectly homogenous, as well as the FS mirrors. A force density $\overrightarrow{a}\cdot \rho_x$ with $|\overrightarrow{a}|=|\overrightarrow{g}|=9.8$ m~s$^{-2}$, is applied on the ULE spacer of density $\rho_s$ = 2210~kg~m$^{-3}$, and on the FS mirrors of density $\rho_m $ = 2203~kg~m$^{-3}$, and it induces a quasi-static elastic deformation, as we study low frequency effects. The simulation provides an optimum solution for which the cavity length variation $k_j$ and residual angle tilt $\kappa_j$ coefficients both vanish, and the sensitivities of these coefficients to changes from the optimum support point positions.

\begin{figure}
\center
  \includegraphics[width=0.5\textwidth]{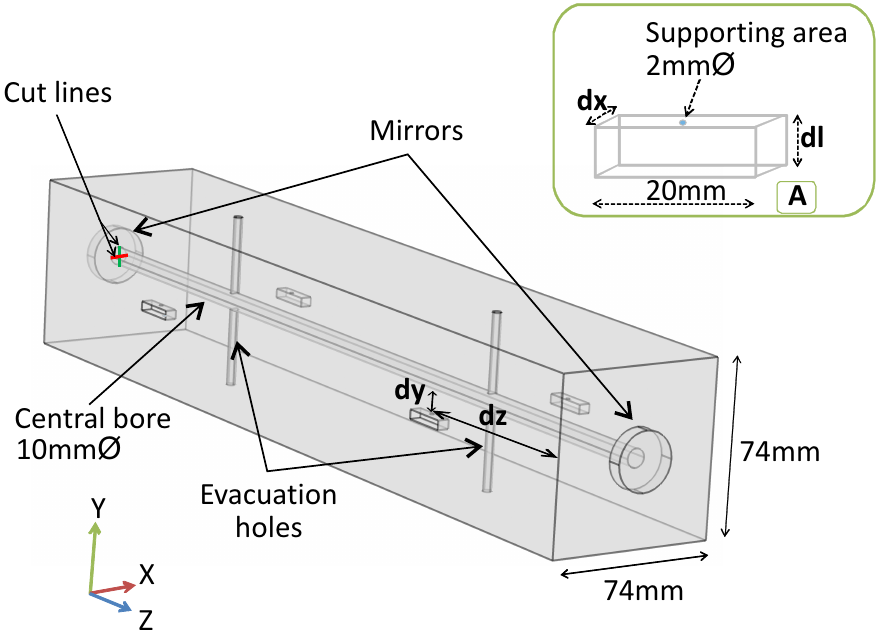}\\
  \caption{Parameters of the $39.5$~cm long cavity. The inset shows the parameters for one of the cut-outs by which the cavity will be supported, and the blue point indicates the fixed contact position of the mechanical support.}\label{Fig_para_cav}
\end{figure}

We consider a cuboid shape of the spacer, which seems to be the preferable choice for long cavities if loaded properly~\cite{tao2007decreased,chen2006vibration,koide2010design}. Four small cut-outs have been designed in the ULE spacer (two on each side) in order to support the cavity. The parameters of the cavity are presented in Fig.~\ref{Fig_para_cav}. All the cut-outs (box (A) in Fig.~\ref{Fig_para_cav}) have the same dimensions. The size of the cut-outs is a compromise between maximum adjustability of the support points and the desire to retain the symmetry between upper and lower half of the spacer by having the support points as close as possible to the central horizontal symmetry plane. The mechanical support of the cavity is in contact with the ULE spacer only in a small circular area of 2~mm diameter to reduce thermal conductivity between the heat shields and the cavity through the supporting legs. The displacement of the mirror and its tilting angle are evaluated from a vertical cut line (for the case of vertical (along $Y$) or axial (along $Z$) accelerations) and a horizontal cut line (for the case of horizontal acceleration (along $X$)) crossing the center of the mirror, as shown by the green and red lines of Fig.~\ref{Fig_para_cav}, respectively.

For finding the optimum mounting points and for estimating the allowed machining tolerances, the position of the entire cut-out within the spacer is varied in the simulations. We first consider vertical acceleration along the $Y$ axis. The volume of the cut-outs and in particular their height $dl$ (dimension of the cut-out along the $Y$ axis) break the symmetry of the cavity with respect to the middle horizontal symmetry plane $(Z,X)$. This results in a finite axial length change under vertical acceleration through Poisson's ratio~\cite{nazarova_vibration-insensitive_2006}. This effect scales with the diameter of the cavity and can be cancelled by optimizing the parameter $dy$ (position of the cut-out along the $Y$ axis, relative to the middle horizontal plane of the cavity).

Mirror tilt is introduced through the asymmetric expansion of the spacer with respect to the middle horizontal plane ($X$,$Z$) due to Poisson's ratio, and through the bending of the cavity around the support points with two extreme cases: supporting in the center leads to downward tilt of the ends of the cavity, whereas supporting at the ends results in an upward tilt. In between these two configurations, there is an optimum position with vanishing total mirror tilt that can be found by adjusting $dz$ (position of the cut-out along the optical axis $Z$, starting from the end faces). The depth of the cut-out $dx$ (from the side face of the spacer towards the optical axis) has a negligible effect on the mirror tilt and length change of the cavity at the optimum parameters for $dz$ and $dy$.

\begin{figure}[h]
 \center
  \includegraphics[width=0.5\textwidth]{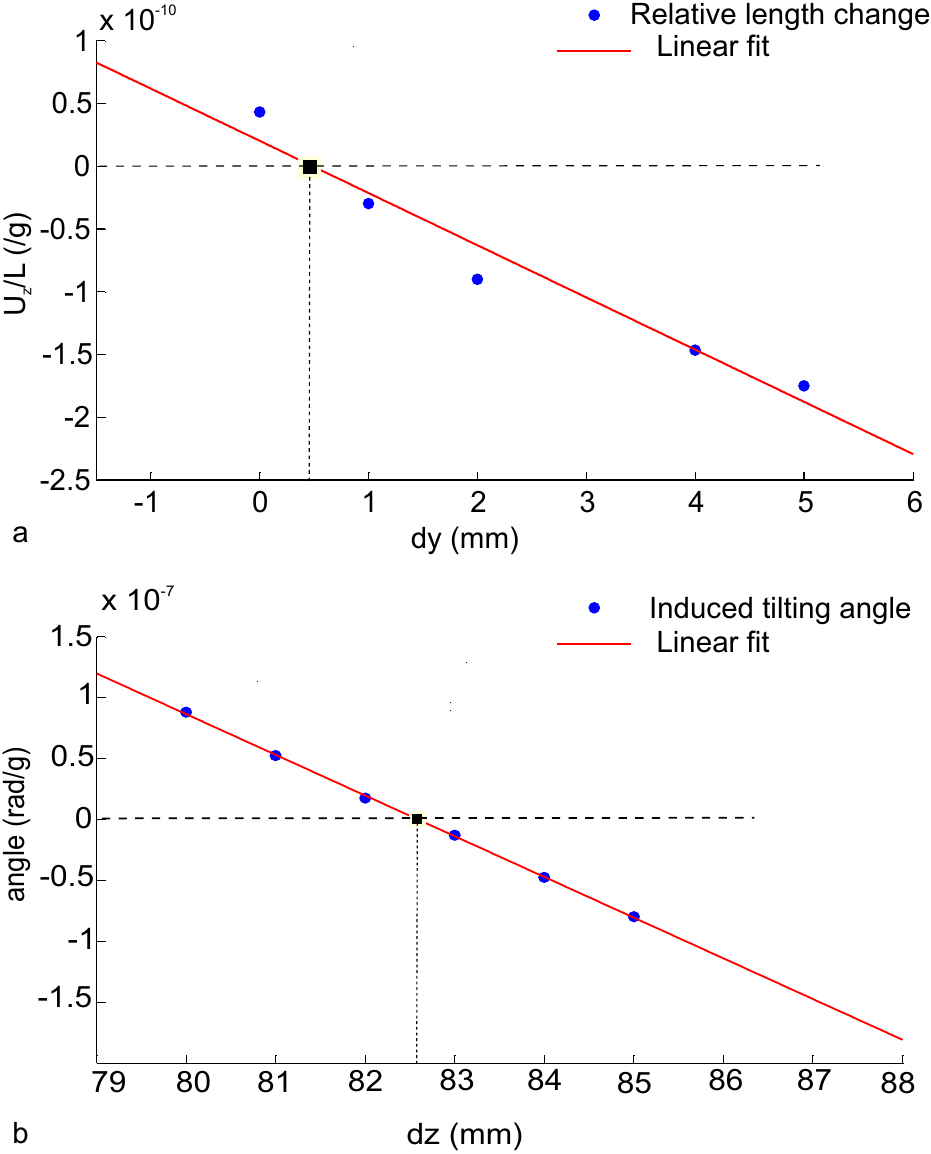}\\
  \caption{Simulated cavity length change and tilt under vertical acceleration as a function of parameters $dy$ and $dz$. (a) Relative cavity length change U$_z/$L at the center of the mirror as a function of the cut-out position $dy$ for fixed $dz=83$~mm. A linear fit provides the optimum value of $dy=0.5$~mm and a sensitivity to machining tolerances of $k_y = 42\times 10^{-12}\cdot \delta dy/(g\cdot mm)$. (b) Cavity mirror tilt as a function of the cut-out position $dz$ for fixed $dy=0.5$~mm. A linear fit provides the optimum value of $dz=82.6$~mm and a sensitivity to machining tolerances in $dz$ of $\kappa_y = 84\times 10^{-12} \cdot \delta dz /(g\cdot mm^2)$.}\label{Fig_dz_dy_sens}
\end{figure}

The relative displacement values at the mirror center as a function of the parameter $dy$ are plotted in Fig.~\ref{Fig_dz_dy_sens}~(a). The tilt angle of the mirror is plotted as a function of the parameter $dz$ in Fig.~\ref{Fig_dz_dy_sens}~(b). The simulations confirm the dependence of the length change and the mirror tilt to the almost independent variables $dy$ and $dz$, respectively. Linear fits to the data provide us with the optimum positions $dz=82.6$~mm and $dy=0.5$~mm. The slope of these fits is a measure for the sensitivity against machining tolerances $\delta dy$ and $\delta dz$ for the cut-outs. For horizontal and axial acceleration, the sensitivity to changes in the support points was determined analogously to the sensitivity in the vertical direction described above.

\begin{table}
\center
  \includegraphics[width=0.5\textwidth]{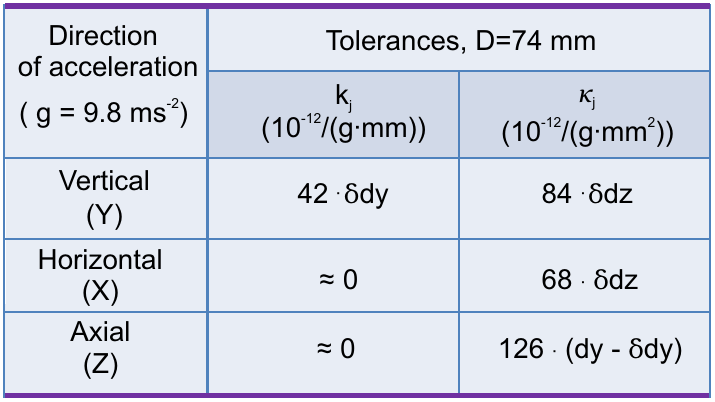}\\
  \caption{Optimum support point positions and acceleration sensitivities to machining tolerances of a 39.5~cm long cavity with a square cross section of 74~mm side length.}\label{Fig_Sens_acceleration}
\end{table}

Table~\ref{Fig_Sens_acceleration} presents a summary of the simulation results as functions of machining tolerances for a 39.5~cm long cavity loaded with vertical, axial, and horizontal acceleration fields. The symmetry of the cavity spacer with respect to horizontal (along $X$) and axial (along $Z$) acceleration is maintained through the mounting structure, thus eliminating length changes along the cavity axis. For the axial acceleration, a mirror tilt occurs due to $dy \neq 0$ and can not be avoided as long as the supporting legs are not exactly on the horizontal symmetry plane ($dy = 0$). By choosing cut-outs of small volume, we reduce the magnitude of the $dy$ parameter to $dy = 0.5$~mm and get a tolerable sensitivity to axial acceleration of $63 \times 10^{-12}/g$ for the ideal case of $\delta dy =0$. For the horizontal acceleration, the optimum cut-out position agrees with the optimum value found for the vertical acceleration within the precision of the simulations.

The measured vibration spectrum in our experimental lab is around $7\mu$g/$\sqrt{Hz}$ at 1~Hz in all three directions. Thus, to achieve a relative length change below the expected thermal noise limit ($\frac{\Delta L}{L}< 10^{-16}$), we will need the machining tolerances for making the supporting cut-outs in the spacer to be in the order of a few $100~\mu$m. The large sensitivity to mirror tilt in long cavities requires an alignment of the optical with the geometrical axis on the mirrors ($\Delta r$) to be precise to within a few $100~\mu$m. Both requirements are achievable with current technology.

\subsection{Influence of the diameter-length ratio}\label{sec:diameter_cavity}

Bending of the mirrors under accelerations orthogonal to the optical axis is caused by a bend of the entire spacer (extremal for support points either in the center or the outer points along the optical axis), and an asymmetric expansion along the cavity axis of the lower and upper half of the cavity due to Poisson's ratio~\cite{nazarova_vibration-insensitive_2006}.
For vertically mounted cavities these two effects can cancel for a certain diameter-length ratio~\cite{millo_ultrastable_2009}.
For the horizontally mounted cavities considered here, the effects add up, and an additional length change is introduced. However, by choosing an appropriate diameter-length ratio, we can balance the sensitivities to length change and tilt.
\begin{figure}
  \includegraphics[width=0.5\textwidth]{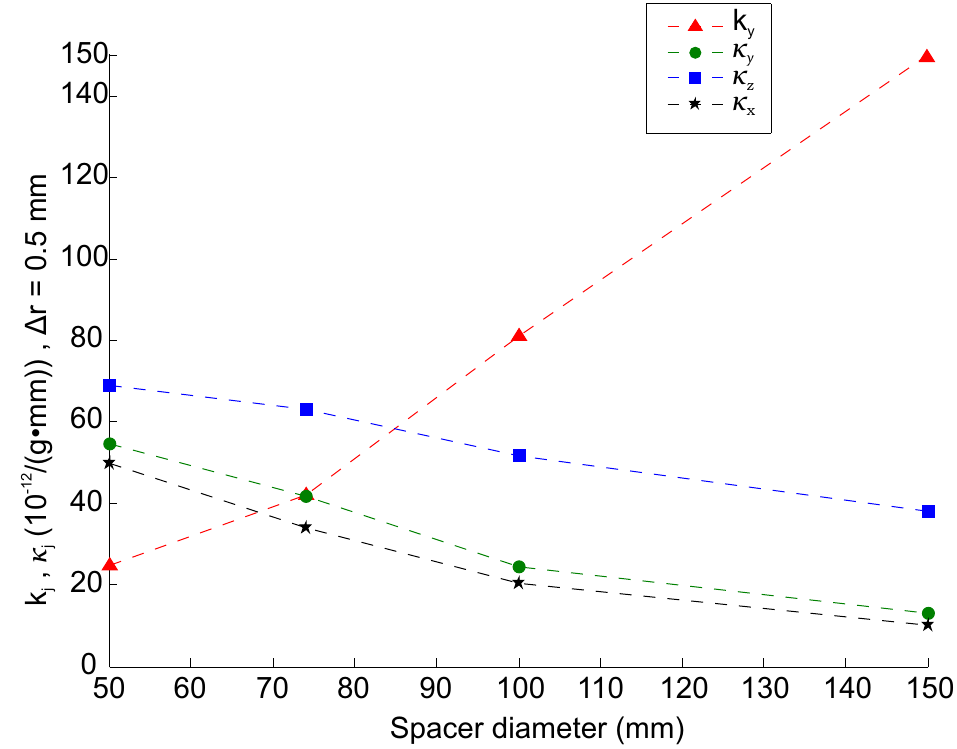}\\
  \caption{Variation of the sensitivity coefficients for mirror tilt $\kappa_j$ and mirror displacement $k_j$ as a function of the spacer cross section, under vertical ($Y$ axis), axial ($Z$ axis), and horizontal ($X$ axis) accelerations. The length of the spacer is fixed to 39.5~cm as well as the shape of supporting cut-outs. We have assumed $\Delta r= 0.5$~mm for the calculation. The dashed lines are a guide to the eye.}\label{Fig_tolerences}
\end{figure}

In Fig.~\ref{Fig_tolerences}, we compare the factors k$_j$ and $\kappa_j$ , under vertical, axial and horizontal accelerations for different square cross sections of a 39.5~cm long cuboid cavity.
The sensitivity to the mirror tilt becomes larger for smaller cross section, whereas the sensitivity to length changes under accelerations perpendicular to the optical axis scales linearly with the cross section~\cite{nazarova_vibration-insensitive_2006}.
We found that, given machining tolerances on the order of a few $100~\mu$m, square cross sections between 65~mm and 85~mm offer a good trade-off between length and tilt sensitivity.

\subsection{Asymmetric spacer and inhomogeneous loading forces}\label{sec:tolerence_cavity}

We investigated the influence of geometrical deviations from the optimum cavity shape to assess the required machining tolerances, which become critical for longer cavities. Through numerical simulations, we estimate the sensitivity of a 39.5~cm long cavity to accelerations in the case of an asymmetric spacer that has an end surface tilted by an angle $\alpha$ around a horizontal axis, maintaining the calculated optimal support points for a perfect rectangular shape. The sensitivity of the cavity becomes $4.9 \times 10^{-9}\alpha/(g \cdot rad)$ and $4 \times 10^{-8}\alpha / (g \cdot rad)$ under vertical and axial accelerations, respectively. The horizontal acceleration is unchanged, since the corresponding symmetry is not broken by the tilt. We expect that a tilt along the vertical axis results in similar sensitivities for horizontal and axial accelerations. We conclude, for a 39.5~cm long cavity of width and depth equal to 74~mm, that the faces of opposing sides of the cavity should be parallel to within 0.1~mrad to achieve a sensitivity of less than $10^{-11}/g$ to accelerations in all directions. This requirement is technically achievable and can probably be further relaxed by experimentally determining new optimal support point positions.

Equal distribution of the total reaction forces $\overrightarrow{F}= m \overrightarrow{a}$ applied between the supporting points is also of crucial importance, and may be a reason for differences between theoretical results and experiential measurements of the vibration sensitivities~\cite{webster_force-insensitive_2011}. An estimate shows that a force difference $\Delta F$ applied between two pairs of supporting legs induces a sensitivity of $7.4 \times 10^{-9}/g \cdot \frac{\Delta F}{F}$ in the axial direction and $2.4 \times 10^{-10}/g \cdot \frac{\Delta F}{F}$ in the vertical direction, for this cavity of mass $m = 4.7$~kg. Thus, equal force distribution within 1$\%$ between the supporting legs is therefore of major importance to achieve a vibration insensitive cavity.
Equalizing the forces applied on the cavity could be either achieved by using elastic supports or by converting the four-point support into an effective three-point support.
\section{Summary \& Conclusion}
We investigated two different approaches to reduce the thermal noise of optical cavities.
The first approach consists of operating the cavity near instability to increase the mode size on the mirrors, thus reducing the dominant source of thermal noise. Whereas the long cavity provides a modest improvement over a comparable cavity operated under stable mode conditions, the improvement for short cavities in terms of stability $\sigma_y$ is better than a factor of 2. Possible issues with the off-resonant coupling of higher-order modes for the near-instability cavities can be solved by a proper choice of the modulation frequency and the optical path length. Furthermore, the mirror alignment requirements are feasible, particularly for the 10~cm short cavity. These results are important for the development of ultra-stable portable lasers, e.g. for portable clocks and future space missions such as LISA, GRACE-fo and STE-QUEST, for which short cavities are advantageous.

The second approach for reducing thermal noise consists of using a long cavity spacer. We presented a mounting design for a 39.5~cm long cavity. This robust design is the result of Finite-Element simulations made for multiple cavity designs that allowed us to find the best diameter-length ratio and machining tolerance for a minimum vibration sensitivity while allowing a large degree of adjustability of the support points. This cavity of reduced theoretical thermal noise limit lower then $10^{-16}$ is now implemented in our laboratory in order to stabilize the interrogation laser of an Aluminum quantum logic optical clock.

\begin{acknowledgements}
This work is supported by the DFG through the Centre for Quantum Engineering and Space-Time Research (QUEST),
by ESA through TRP AO4640/05/NL/PM and GSTP AO/1-6530/10/ NL/NA, and by the European Metrology Research Programme (EMRP). JRP SIB04. J.B.W. acknowledges support from the Hannover School for Laser,
Optics and Space-Time Research (HALOSTAR) and the German National Academic Foundation (Studienstiftung des deutschen Volkes).
\end{acknowledgements}


\end{document}